\documentstyle[11pt,epsfig]{article}
\def\be{\begin{eqnarray}}\def\ba{\begin{eqnarray}}
\def\ee{\end{eqnarray}}\def\ea{\end{eqnarray}}
\def\ben{\begin{enumerate}}\def\bitem{\begin{itemize}}
\def\een{\end{enumerate}}\def\eitem{\end{itemize}}
\def\no{\nonumber\\}
\def\ie{{\it i.e}}\def\bi{\bibitem}

\def\Kp{$K^+$}\def\Km{$K^-$}

\def\prl{Phys. Rev. Lett.}

\def\Tr{{\mbox{Tr}}}

\def\roughly#1{\mathrel{\raise.3ex\hbox{$#1$\kern-.75em%
\lower1ex\hbox{$\sim$}}}}

\def\A0{A_0}
\def\bq{\begin{equation}}
\def\eq{\end{equation}}
\def\la{\langle}\def\ra{\rangle}
\def\Km{K^-}\def\Kp{K^+}\def\K0{K^0}

\begin{document}
\begin{titlepage}
\begin{center}

 \vskip 1.5cm
{\Large \bf Kaon Condensation and Enforced Charge Neutrality \\
of the Color-Flavor Locked Phase }

\vskip 2.3cm
    {\large Youngman Kim$^{(a)}$, Dong-Pil Min$^{(a)}$ and Mannque  Rho$^{(b,c)}$ }
\vskip 0.5cm

(a)   {\it  Department of Phyiscs and Center for Theoretical Physics, Seoul National University,
Seoul 151-742, Korea}

(b)  {\it Service de Physique Th\'eorique, CEA Saclay, 91191
Gif-sur-Yvette  Cedex, France}

(c)  {\it School of Physcis, Korea Institute for Advanced Study,
Seoul 130-012, Korea}

\end{center}

\vskip 1cm

\centerline{(\today) }
 \vskip 1cm

\centerline{\bf Abstract}
 \vskip 0.5cm

We investigate the influence of charged kaon condensation in
hadronic sector on the formation of the electron-free charge
neutral quark matter of equipartition of three light flavors,
i.e., a phenomenon called ``enforced electrical neutrality" of the
color-flavor locked (CFL) phase. We employ the chiral quark model
which is assumed to be applicable, ``bottom-up," toward chiral
restoration while accessible to kaon condensation in the hadronic
sector. Approaching ``bottom-up" the high density regime is more
appropriate in addressing the properties of compact stars than
approaching ``top-down" from asymptotic QCD. It turns out that the
presence of hadronic kaon condensation makes the electron-free
charge-neutral CFL phase energetically more favorable with a
relatively  small color-superconducting gap $\Delta_0\sim 20$ MeV
compared to $\Delta_0\sim 70$ MeV required in the absence of such
kaon condensation.

\end{titlepage}
\newpage

\section{Introduction}
The phenomenon of color superconductivity in high density quark
matter at relatively low temperature along with its physical
consequences has recently been extensively investigated. For
recent reviews, see \cite{review}. It turns out that in high
density quark matter with three flavors, the color-flavor-locked
(CFL) phase is energetically favored, exhibiting a variety of
interesting physics such as enforced electric charge
neutrality\cite{rawi} and kaon condensation~\cite{ts}.

Kaon condensation in the CFL phase (that we shall refer to as
``CFL kaon condensation") was first investigated by Sch\"afer
\cite{ts} who showed that negative kaons can be condensed in
high-density quark matter. The principal mechanism for this
phenomenon is that in the CFL phase, the spectrum of Goldstone
bosons is inverted so that the kaons turn out to be the lightest
modes. When the electron chemical potential $\mu_e$ is equal to
the kaon mass $m_K$, the electrons turn into kaons which then
condense. In the Sch\"afer process, the presence of electrons is
essential. Although this resembles what happens in kaon
condensation that takes place in hadronic phase~\cite{kn,bkrt}
(referred to as ``hadronic kaon condensation" in what follows) in
that it is the kaon mass that plays the key role, there is a
crucial difference. In hadronic kaon condensation, it is the
conspiracy between matter density and chiral symmetry-breaking
(strange quark mass) effect that triggers the phase change, so the
larger the symmetry breaking, the earlier the phase change. The
CFL kaon condensation, in contrast, exploits the smallness of
non-zero symmetry breaking.

The effects of non-zero electron chemical potential and a finite
strange quark mass on CFL matter have been studied in \cite{bs}.
It has been shown there that kaon or pion condensation takes place
as a response to an external {\it stress} generated by quark
masses. In particular, it was argued that if the CFL matter were
in contact with a hadronic matter that supports a large electron
chemical potential, the surface layer would likely be $K^-$- or
$\pi^-$-condensed. The reason why the electron chemical potential
induces $K^-$ or $\pi^-$ condensation is that a positive electron
chemical potential lowers the energy of negatively charged
Goldstone modes in the CFL phase~\cite{bs},
 \ba
E_{\pi^\pm}=\pm\mu_e+m_{\pi^\pm}
,~~E_{K^\pm}=\pm\mu_e+m_{K^\pm}.\nonumber
 \ea
A recent study, which {\it does not include} the effects of the
large electron chemical potential, shows that with nonzero quark
masses, the most relevant phases in nature are $CFLK^0$ (3-flavor,
color superconducting and $K^0$ condensed phase), $CFLK^-$ and/or
$CFLK^+$ depending on densities~\cite{ksr} rather than the
symmetric charge neutral CFL phase with equal numbers of $u$, $d$
and $s$ quarks~\cite{rawi}. The existence of meta-stable
non-topological domain walls associated with broken $U(1)_Y$ in
high density quark matter with $K^0$ condensation has also been
investigated by Son~\cite{son}. Kaplan and Reddy~\cite{krv}
discussed the existence of stable vortices excited by the
spontaneous breaking of $U(1)_Y$ and $U(1)_{EM}$ in $K^0$ and
$K^+$ condensations respectively.

The ultimate goal of research in the physics of high density
matter is to understand the formation and structure of the
interior of compact stars which involve supernova explosion and
collapse into dense compact stars. In this process, nature follows
the route from low density to high density. The studies cited
above all approach the density regime of interest coming down
(i.e., ``top-down") from asymptotic density at which QCD provides
a controllable tool. Whether or not one can actually access,
top-down, with sufficient accuracy the relevant density regime is
not yet clear. What we propose to do in this paper is to take the
other route, namely, start with the low-density regime which is,
though theoretically un-controlled, phenomenologically understood
and access bottom-up the relevant density regime. This approach
unfortunately suffers from lack of reliable theoretical tools,
with QCD being untractable in the nonperturbative regime, so we
are forced to rely on models. This paper is an initial attempt to
make a bridge from lower density to what can happen at higher
density.

At relatively low density, hadrons are the relevant degrees of
freedom. Kaon condensations in terms of hadronic variables have
been studied extensively since mid 1980's following the seminal
work of Kaplan and Nelson~\cite{kn}. Kaplan and Nelson showed in
tree order with $SU(3)_L\times SU(3)_R$ chiral Lagrangian that
kaons could condense at a density around $3\rho_0$. Subsequently a
new mechanism for kaon condensations going beyond the tree order
which is consistent with kaon-nuclear interactions was proposed by
Brown, Kubodera, Rho and Thorsson~\cite{bkrt}. This work as well
as others that followed it confirmed that the critical density
lies in the range $2\rho_0 <\rho_c < 4\rho_0$, more or less in
accord with the first prediction of \cite{kn}.  For reviews see
\cite{chlee}. At densities higher than that of normal nuclear
matter, hyperons or more generally strange-quark degrees of
freedom -- that we shall refer to as strange matter -- can become
relevant. The effect of hyperons or of strange matter has been
studied by several authors. It is found that the presence of
strangeness in matter tends to push hadronic kaon condensation to
higher densities or depending on parameters -- which are hard to
pin down -- even out of relevant density regime. This aspect is
reviewed in \cite{hyper}.

The goal of the present work is to study how $\Km$ condensation
aided by electrons that can take place in hadronic sector affects
the formation of color superconductivity (CSC) in quark sector.
Here we are imagining dialing the density from the regime where
hadronic kaon condensation is present to the regime where CSC in
the CFL phase is realized. In describing CSC in dense quark
matter, one assumes, to start with, that the normal phase of the
quark matter is in the form of a Landau Fermi liquid, stable
against quark($q$)-anti-quark($\bar{q}$) pair condensation but
unstable against diquark condensation. What we wish to do here is
to start from the density regime in which $q\bar{q}$ pairs are
condensed, so that Goldstone bosons of the $\bar{q}q$ type are
present and climb up in density to the regime where CSC is
presumed to take place. Thus we are approaching Cooper pairing
from a sort of non-Fermi-liquid ground state. As a first step to
study CSC with such nontrivial vacuum configurations, we consider
quark matter in the presence of $\Km$ condensation. To do this, we
take the chiral quark model~\cite{mg} which is believed to be
relevant near the chiral phase
transition~\cite{riska-brown,BR96,BR01}. We first study hadronic
$\Km$ condensation with this model and assure that it reproduce
more or less what has been obtained in chiral perturbation theory
in the hadronic sector. We then extrapolate this model to the
density regime of CFL phase and examine the consequence on, among
others, the enforced electric charge neutrality nature of the CFL
phase. What we are doing can be understood in terms of the NJL
(Nambu-Jona-Lasinio) model. In the NJL model, one considers $\la
\bar q q\ra$ and $\la q q\ra$ condensations, and finds that the
phase with $\la \bar q q\ra$ condensate is energetically favored
in the low density regime while the other one with $\la q q\ra$
condensate is at high density. We find that this is true whether a
mixed phase exists~\cite{rssv} or not~\cite{ha}. In fact,  the
presence of a mixed phase is not crucial.

The structure of this paper is as follows. In section 2, we
introduce the chiral quark model and study kaon condensations in
the frame work of the model with and without hyperons or strange
matter. We investigate the effect of kaon condensation on the
enforced charge neutrality of the CFL phase in section 3. We
summarize our results in section 4.

\section{Kaon Condensation in the Chiral Quark Model}

In this section, we describe hadronic kaon condensation in the
context of the chiral quark model and compare our results with
those obtained in heavy-baryon chiral perturbation theory
(HBChPT). To do this, we calculate the in-medium $K^-$ mass using
the chiral quark model defined by the Lagrangian
 \ba
{\mathcal L}&=&{\bar\psi}(i {\not\! D}+{\not\! V})\psi +
g_{A}{\bar \psi}{\not\! A}\gamma_5\psi
               - {\rm  m} {\bar \psi}\psi \no
               &&+ \frac{1}{4}f_\pi^2
\Tr(\partial^{\mu}\Sigma^{\dag}\partial_{\mu}\Sigma)
               - \frac{1}{2}\Tr(F^{\mu\nu}F_{\mu\nu}) + \ldots
\label{lag}
\ea
where
\ba
\psi &=&
 \left( \begin{array}{c}
u\\
d\\
s
 \end{array} \right),\\
D_\mu&=&\partial_\mu+i g G_\mu,~~~ G_\mu=G_\mu^a T^a, \no
V_\mu&=&{i \over
2}(\xi^{\dagger}\partial_\mu\xi+\xi\partial_\mu\xi^{\dagger}),\no
A_\mu&=&{i \over
2}(\xi^{\dagger}\partial_\mu\xi-\xi\partial_\mu\xi^{\dagger}),\no
\xi&=&e^{(i \Pi/f_\pi)},~~~ f_\pi\simeq 93~ MeV,~
\Sigma=\xi\xi\nonumber
\ea
and
\ba
\Pi={1 \over\sqrt{2}}\left( \begin{array}{ccc}
                            \sqrt{1 \over 2}\pi^0+\sqrt{1 \over 6}\eta&
\pi^+& K^+ \\
                            \pi^-&-\sqrt{1\over 2}\pi^0+\sqrt{1\over
6}\eta& K^0 \\
                            K^- & \bar{K}^0& -{2 \over \sqrt{6}}\eta
                            \end{array} \right).\label{L1}
\ea Here, $\rm m$ denotes the constituent quark mass generated by
spontaneous chiral symmetry breaking and is approximately
$350MeV$. To investigate kaon condensation, we need the
symmetry-breaking term
 \ba
{\mathcal L_M}= -\frac{1}{2}c_1\bar\psi(\xi^\dagger {\mathcal
M}\xi^\dagger +\xi {\mathcal M} \xi)\psi \label{L2} \ea where
$c_1\approx 1$ \cite{georgi} and \ba \left(
\begin{array}{ccc}
 m_u& 0& 0 \\
0& m_d& 0 \\
0& 0 & m_s
    \end{array} \right).\nonumber
\ea In the chiral quark model, one has the advantage of a
systematic chiral power counting (as summarized in Appendix) to
account for the interactions of constituent quarks with Goldstone
bosons and of weak-coupling expansion for interactions with gluons
(with $\alpha_s\approx 0.23$)~\cite{mg}. Within the scheme, there
are no free parameters to the order we are considering in contrast
to the plethora of uncertainties in coupling constants present in
the models involving hyperons~\cite{hyper}.

\begin{figure}
\centerline{\epsfig{file=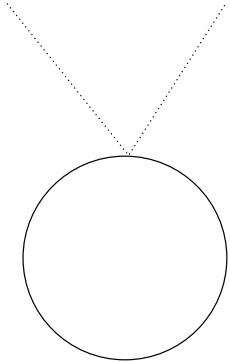,width=3.0cm}} \caption{\small
The kaon self-energy. Dotted lines denote kaons and solid ones are
for quarks within the quark Fermi sea.}\label{kaon}
\end{figure}

We begin with the relevant part of the Lagrangian for the problem
at hand which is of the form
 \ba {\mathcal L}_K&=&
\frac{i}{4f_\pi^2}[\bar u (K^+\not\!\partial K^- -\not\!\partial
K^+K^- )u +
\bar s (K^-\not\!\partial K^+ -\not\!\partial K^-K^+ )s ]\label{LL1}\\
&+& \frac{1}{2f_\pi^2}(m_u +m_s) [\bar u K^+K^- u+ \bar s K^- K^+
s].\label{LL2} \ea The lowest order self-energy correction with
this Lagrangian comes from the graph shown in Fig. \ref{kaon}.

Before going into detailed calculations, one can see qualitatively
what can happen with and without strange matter. By ``strange
matter," we mean, within the framework of chiral quark model,
non-zero ground state expectation value of the strange quark pair,
$\la\bar{s}s\ra_F$. This will be zero if the strange quark Fermi
sea is empty. Hyperon matter in terms of hadronic degrees of
freedom is replaced in this scheme by a strange-quark Fermi sea.
One can see from (\ref{LL1}) that {\it $u$ quarks decrease
(increase) the in-medium mass of $K^-$ ($K^+$), while  $s$ quarks
decrease (increase) the in-medium mass of $K^+$ ($K^-$). Therefore
we expect that at some high densities, the in-medium $K^+$ and
$K^-$ masses can become degenerate, $\ie$ $m_{K^+}^* =m_{K^-}^*$}.
This interesting result seems, however, at odds with, for example,
the results in Ref. \cite {bkrt} in which $m_{K^+}^*$
($m_{K^-}^*$) is found to increase (decrease) with increasing
density. This discrepancy can be understood by noting that in Ref.
\cite {bkrt}, hyperons (or strange matter) are absent. One can
understand simply why in the absence of strange-quark Fermi sea, a
$\Km$ gets lighter while a $\Kp$ gets heavier. In the chiral quark
picture, $K^-\approx \bar us$ and $K^+\approx u\bar s$, so we
expect that putting a $K^+$ on top of the Fermi seas of $u$ and
$d$ quarks is suppressed by the $u$ quark in the Fermi sea. Now
let us increase the baryon density and see what happens there. As
density increases, strange quarks start making a strange Fermi sea
and eventually at very high density, we can have a Fermi sea
composed of equal numbers of $u$, $d$ and $s$ quarks. In terms of
the quark picture given above, it is reasonable to expect that the
masses of $\Km$ and $\Kp$ become equal at very high density. Thus
the CFL picture arises naturally in this picture.

\subsection{Kaon condensation without strange matter}

Consider kaons in a medium that consists only of up and down quark
Fermi seas without strange Fermi sea. We believe this system to be
analogous to hadronic matter without hyperons. To do this, we
treat the quarks as ``heavy" and use the heavy-fermion
approximation as in HBChPT. The kaon self-energy computed with
Fig. \ref{kaon} in this approximation is
 \ba -i\Sigma_K (q_0) = i
[\frac{3}{4} \frac{(m_u+m_s)}{f_\pi^2} +\frac{3}{4}
\frac{q_0}{f_\pi^2}]\rho_B
 \ea
where $\rho_B$ is the baryon density $\rho_B=\frac{2}{3}\rho_Q$
with $\rho_Q$ being quark density~\footnote{In general, we have
$\rho_B=(\rho_u+\rho_d+\rho_s)/3$. In case of the symmetric
nuclear matter characterized by $\rho_u=\rho_d\equiv\rho_Q$ and
$\rho_s=0$, we get $\rho_Q=(3/2)\rho_B$.}. To obtain the in-medium
kaon mass, we need to solve the dispersion equation
 \ba
m_K^{\star 2} =m_K^2 + \Sigma_K (q_0=m_K^{\star }).
 \ea
Defining $x=\frac{m_K^\star}{m_K}$ and taking $m_u\approx 6$ MeV,
$m_s\approx 240$ MeV as in \cite{chlee} and $m_K\approx 500$ MeV,
we rewrite the dispersion equation \ba x^2 +0.24cx +0.12c-1 =0 \ea
where $c=\rho_B/\rho_0$, {\it i.e}, the ratio of baryon density to
normal nuclear matter density. For instance taking c=1,
$\rho_B=\rho_0$, we get
 \ba
m_{K^-}^\star\approx 410\  {\rm MeV}~,~~ m_{K^+}^\star\approx 533\
{\rm MeV}.
 \ea
In HBChPT, one obtains $m_{K^-}^\star\approx 360$
MeV~\cite{chlee}. At higher densities, we get
 \ba c=2&:&m_{K^-}^\star\approx 330\  {\rm MeV} \no
c=3&:&m_{K^-}^\star\approx 260\  {\rm MeV}\no
c=4&:&m_{K^-}^\star\approx 193\  {\rm MeV}.
 \ea
Noting that for densities of $(2\sim 4)\rho_0$, $200\ {\rm MeV}
<\mu_e <300\ {\rm MeV}$, we expect kaon condensation to take place
around $\rho\approx 3\rho_0$.

We consider these results to be unreliable for two reasons. First
we are ignoring strange quark matter which is unjustified within
the model. Second the parameters in the Lagrangian must be
density-dependent as recently shown in a different
context~\cite{HKR}. We shall consider the former in the next
subsection. Here we take into account the density dependence of
the parameters, we adopt the BR scaling approach~\cite{br} and use
$f_\pi^\star =(1-0.15\rho/\rho_0 )f_\pi$ to compare with the
results in Ref. \cite{brkc}. The results we get are
 \ba
c=1&:&m_{K^-}^\star\approx 380\ {\rm MeV} \no
c=2&:&m_{K^-}^\star\approx 190\ {\rm MeV}\nonumber
 \ea
The effective chiral Lagrangian with BR scaling used in Ref.
\cite{brkc} gives $m_{K^-}^\star\approx 330\ {\rm MeV}$ in
symmetric nuclear matter at $\rho=\rho_0$.

In summary, we find that without strange quark matter, kaon
condensation occurs roughly at about $\rho=3\rho_0$ independently
of whether BR scaling is incorporated or not.

\subsection{Kaon condensation with strange matter}
It has been argued that hyperons appear around
$(2-4)\rho_0$~\cite{hyper}, the range of density close to the
critical density for kaon condensation. We study in this
subsection the effect of strangeness presence. Here for
simplicity, we shall take hyperon or strange quark population as
an {\it external parameter} to be varied.

\begin{figure}
\centerline{\epsfig{file=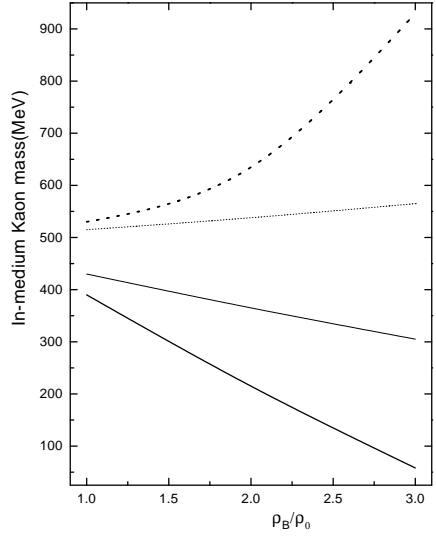,width=6.3cm}} \caption{\small
In-medium kaon mass vs baryon density: the dotted lines are for
$m_{K^+}$ and the solid lines for $m_{K^-}$, thin lines without BR
scaling and thick lines with BR scaling, for $y=0.2$. }\label{y02}
\end{figure}

\begin{figure}
\centerline{\epsfig{file=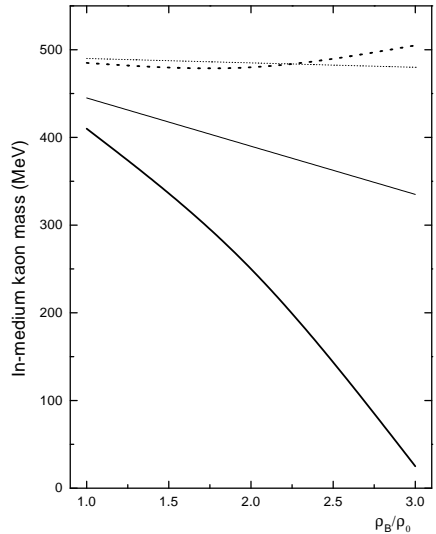,width=6.3cm}} \caption{\small
In-medium kaon mass vs. baryon density: The dotted lines are for
$m_{K^+}$ and the solid lines for $m_{K^-}$, without BR scaling
(thin lines) and with BR scaling (thick lines), for $y=0.5$.
}\label{y05}
\end{figure}

To describe kaon condensation in the presence of strangeness which
is relevant in higher densities than without, we have to treat the
quarks relativistically since the constituent quark mass must
drop~\cite{HKR}. The relativistic self-energy of kaon of Fig.
\ref{kaon} reads
 \ba
\Sigma_K q_0) & = &-\frac{(m_u+m_s)}{2f_\pi^2}
[3M_uI(q_F^u,M_u)+3M_sI(q_F^s,M_s)]\no &&-
\frac{q_0}{2f_\pi^2}[\rho_u -\rho_s]\label{is-1} \ea where \ba
I(q_F, M)&=&\frac{1}{2\pi^2}[q_F\sqrt{q_F^2 + M^{2}}
 - M^{2} ln ( \frac{q_F + \sqrt{q_F^2+ M^{2}}}{M})],\no
\rho_q&=&\frac{1}{\pi^2}q_F^3.
 \ea
Here $m_q$ is the current quark mass and  $M_q$ the constituent
quark mass. It is immediately clear from the second line of
(\ref{is-1}) that the presence of strange matter will increase
$K^-$ mass. However, {\it the negatively charged kaon mass will
drop with density as long as $\rho_u>\rho_s$}. It is convenient to
define $y$ as
 \ba
\rho_B&=&\frac{1}{3}(\rho_u+\rho_d+\rho_s)\no
&=&\frac{1}{3}(1+1+y)\rho_Q
 \ea
where $\rho_u=\rho_d=\rho_Q$ and $\rho_s=y\rho_Q$. The in-medium
kaon mass is obtained from the solution of the dispersion relation
 \ba
D_K^{-1}(\omega )&=&\omega^2-m_K^2-\Sigma_\rho (\omega )\no &=& 0,
\ \ {\rm with} \ \ \omega=m_K^\star,\label{de1}
 \ea
where $m_K^\star$ denotes the in-medium kaon mass. Taking  $m_u=6$
MeV, $m_d=12$ MeV, and $m_s=240$ MeV as in \cite{kn}, we obtain
the results given in  Fig. \ref{y02}, Fig. \ref{y05} and Table 1.
As expected, the mass of $K^-$ (thin solid line) decreases with
density. However, it decreases too slowly, thereby shifting the
onset for kaon condensation to a much higher density, $\rho>
8\rho_0$ or even excluding it. This is essentially the result
found when hyperons are introduced in the conventional way in a
hadronic field theory~\cite{hyper}. In medium, however, this
treatment is incomplete since it ignores the ``intrinsic" density
dependence of the parameters of the Lagrangian required by
matching to QCD as described in \cite{HKR,HS}.  In specific terms,
this means that we have to take into account that masses and
coupling constants must scale with density. We do not know how to
compute the density dependence from first principles, so to
proceed, we resort to BR scaling~\cite{br} and take
 \ba
\frac{f_\pi^\star}{f_\pi}=\frac{M_q^\star}{M_q}=\frac{1}{1+0.28\rho_B/\rho_0}.
 \ea
Inserting this into the dispersion equation (\ref{de1}), we
obtain the results summarized in  Fig. \ref{y02}, Fig. \ref{y05}
and Table 1.

  \begin{center}
  Table 1 : \parbox[t]{5.3in}{In-medium kaon mass without BR scaling (second  column)
and with BR scaling (third column) when $y=1$. }
  \end{center}
  $$
  \begin{array}{|r||r|r|}
  \hline
   c &m_{K^+}^\star = m_{K^-}^\star (MeV) & m_{K^+}^\star = m_{K^-}^\star (MeV)\\
  \hline
  \hline
  1  & 463    & 440 \\ \hline
  2  & 426  & 320 \\ \hline
  2.84  & 394  & 26\\ \hline
  5  & 300  & .\\ \hline
  8.3  & 30 & .\\ \hline
  \end{array}
  $$
  \vskip 0.3cm
In Table 1, we consider a quark matter composed of equal numbers
of $u$, $d$ and $s$ quarks and find that $K^+$ and $K^-$ masses
become equal as expected. The electron chemical potential is known
to be lowered substantially when hyperons or strange quarks are
present in matter. For instance, it can be as low as $\sim 150$
MeV for $\rho_0\le\rho_B\le 6\rho_0$~\cite{glen}. For a given
baryon density and a lepton number per baryon, the Fermi momenta
of $u$, $d$, $s$, $\nu$ and electron are determined by the
constraints of charge neutrality and $\beta$
 equilibrium~\cite{iwa} from which the electron Fermi momentum or
the electron chemical potential is found to be $108\ {\rm MeV}
<\mu_e<223\ {\rm MeV}$ depending on the value of the lepton number
per baryon at $\rho=\rho_0$.

As one can see in Fig. \ref{y02} and Fig. \ref{y05}, with BR
scaling, the in-medium kaon masses are $\sim 100$ MeV around
$\rho\sim 3\rho_0$. Thus we arrive at the qualitative conclusion
that kaon condensation can take place around $\rho\sim 3\rho_0$
regardless of the presence or absence of strange matter.

\section{Kaon Condensation and Color-Flavor Locking}
In this section, we turn to the question of whether and how the
presence of hadronic kaon condensation discussed above influences
the enforced charge neutrality of the CFL phase. Note that here we
are not talking about kaon condensation in the CFL phase. As
discussed above, we expect $K^-$ condensation in the vicinity of
$\rho\sim 3\rho_0$ even with strange matter when BR scaling is
incorporated. Therefore we start by assuming that $\Km$ condenses
in dense neutron-star medium and write
 \ba
\la K^- \ra = v\exp^{-i\mu_K t}.
 \ea
Given this condensation, the interaction terms in (\ref{LL1}) will
change the chemical potential of $u$ and $s$ quarks while
(\ref{LL2}) will reduce quark masses. Focusing on the change in
the chemical potentials, we write
 \ba
{\cal L}=\bar\psi (i\not\!
\partial +\bar\mu\gamma_0-\mu_e Q\gamma_0)\psi +\delta\mu_K(\bar
u\gamma_0 u -\bar s\gamma_0 s)\label{kkk}
 \ea
where $\bar\mu$ is the baryon chemical potential, $\mu_e$ the
chemical potential for electric charge and $\delta\mu_K$ denotes a
correction from kaon condensation,
 \ba
\delta\mu_K\equiv \frac{v^2}{2f_\pi^2}\mu_K .\nonumber
 \ea
{}From (\ref{kkk}), we observe
 \ba
\mu_u&=&\bar\mu -\frac{2}{3}\mu_e +\delta\mu_K\no \mu_s&=&\bar\mu
+\frac{1}{3}\mu_e -\delta\mu_K.
 \ea
We note here that kaon condensation decreases the strange-quark
chemical potential $\mu_s$ since it provides the system with an
economic way of piling up {\it strange} quarks without increasing
their Fermi momenta.

In order to import the above observation into the CFL phase, we
briefly summarize the results of Ref. \cite{rawi}. It was shown in
Ref. \cite{rawi} that quark matter in the color-flavor locked
phase is automatically charge neutral and no electrons are
required. To demonstrate this, they perform a simple calculation
with two flavors called $1$ and $2$ and assume $m_1=0$ and
$m_2=m_s\neq 0$. The chemical potentials of the quarks are \ba
\mu_1&=&\bar\mu-\delta\mu\no \mu_2&=&\bar\mu+\delta\mu \ea where
$\mu_e=2\delta\mu$. Here we can identify the quark $1$ with an
$up$ quark and $2$ with a $strange$ quark. Comparing the free
energy of the BCS state and that of the unpaired state, they
obtain
 \ba
\Omega_{BCS}-\Omega_{normal}=\frac{\bar\mu^2}{\pi^2}
[(\frac{m_s^2}{4\bar\mu}-\delta\mu )^2-\frac{\Delta_0^2}{2} ].
 \ea
Then it is easy to see that the BCS state is the global minimum of
the system if
 \ba
|\frac{m_s^2}{4\bar\mu}-\delta\mu |<
\frac{\Delta_0}{\sqrt{2}}.\label{con}
 \ea

   \begin{center}
    Table 2 : \parbox[t]{5.3in}{The chemical potential difference $\delta\mu_t$ ($MeV$)
as a function of baryon density with different sets of parameters
with $\mu_e=150$ MeV. Here I, II and III denote the values taken
from Tables 3, 4 and 5 of Ref. \cite{tml} respectively. The
rotation angle (in degrees) is defined by $\theta\equiv
(\sqrt{2}v)/f_\pi$.}
  \end{center}
  $$
  \begin{array}{|r||r|r||r|r|r||r|r|r|}
 \hline
I& & &II& & & III& &\\ \hline
   \rho_B & \theta_{min} & \delta\mu_t
& \rho_B & \theta_{min}  & \delta\mu_t
& \rho_B & \theta_{min}  & \delta\mu_t \\
  \hline
  4.18  & 0    & 75 & 3.08 & 0 & 75 & 2.42 & 0  & 75 \\ \hline
  4.68  & 28.9  & 66 & 3.58 & 39.3 & 57 & 3.42 & 80.5 & 0.13\\ \hline
  5.18  & 40.6  & 56 & 4.58 &67.3 & 23 & 4.42 & 98.8 & -37 \\ \hline
  5.68  & 48.9  & 48 & 5.58 & 80 & 2 & 5.42 & 105.7 & -53 \\ \hline
  \end{array}
  $$

\vskip 0.3cm

In Table 2, we display the values of $\delta\mu_t$ for different
sets of parameters given in Ref. \cite{tml}. The results in Table
2 indicate that kaon condensations could facilitate electric
charge neutrality of CFL phase within a wide range of parameter
choice.

From the development made so far, it is clear what we should do to
see the effect of hadronic kaon condensation~\footnote{More
properly, we should compare the free energy of the BCS state with
that of the state with kaon condensation, but in this work we are
looking at the effects of kaon condensation only to the change of
quark chemical potentials. Comparing the ground state energy of
the CFL phase with that of the phase with kaon condensation is
being investigated and will be reported elsewhere.}. The key point
is that with kaon condensation, the condition (\ref{con}) is
modified to
 \ba
|\frac{m_s^2}{4\bar\mu}-\delta\mu_t |<
\frac{\Delta_0}{\sqrt{2}}.\label{con-r}
 \ea
where $\delta\mu_t=\delta\mu -\delta\mu_K$. Since $\mu_e=\mu_K$,
we can rewrite \ba
\delta\mu_t=\frac{1}{2}\mu_e(1-\frac{v^2}{f_\pi^2}).
 \ea
Now to see the effect of kaon condensation, we take the example of
Ref.\cite{rawi}. With $m_s=200$ MeV, $\bar\mu=400$ MeV and
$\mu_e=2\delta\mu=150$ MeV, the authors of \cite{rawi} obtained
$\Delta_0 > 50$ MeV from (\ref{con}). With kaon condensation, for
instance with $\theta_{min} \equiv\frac{\sqrt{2}v}{f_\pi}\approx
75$ (in degrees) from Ref. \cite{tml}, we obtain
$\Delta_0/\sqrt{2}> |25-11|$ and therefore $\Delta_0 > ~20$ MeV.
We take this to imply that with kaon condensation, the enforced
charge neutrality of the CFL phase is energetically more favored
with relatively smaller color superconducting gap compared to the
case without kaon condensations.

\section{Summary}
In this work, we studied, bottom-up, charged kaon condensation
aided by electrons within the framework of chiral quark model and
their possible effects on the charge neutral CFL phase. When a
quark matter is in contact with a hadronic matter, a large
electron chemical potential can be supported by the hadronic
matter. We have shown that kaon masses drop generically to $\sim
100$ MeV in the vicinity of $\rho=3\rho_0$ when BR scaling is
incorporated, and hence an electron chemical potential around
$\mu_e\sim 100$ MeV will be able to trigger electron-aided kaon
condensation. In the presence of hyperons or strange quarks, the
electron chemical potential comes out to be $\sim 150 MeV$ for
$\rho_0\le\rho_B\le 6\rho_0$~\cite{glen,iwa}. Consequently kaon
condensations can take place at around $\rho=3\rho_0$ whether
strange matter is present or not. Extrapolating to higher density,
it is argued that kaon condensation as formulated in the model can
render the enforced charge neutrality of the CFL phase
energetically more favorable with relatively small color
superconducting gap than without kaon condensation. This argument
is not without caveats, however, and in order to make it firmer,
it will be necessary to compare the ground state energy of the CFL
phase with that of the phase with kaon condensation. This issue is
under investigation.

\vskip 1cm
\subsection*{Acknowledgments}

We are grateful for useful discussions with Deog-Ki Hong. 
This work is supported partly by the BK21 project of the Ministry of Education,
 KOSEF Grant 1999-2-111-005-5
  and KRF Grant 2001-015-DP0085. The work of MR was initiated at SNU supported 
 by the Brain Korea 21 and
completed at KIAS.

\newpage

\section*{Appendix A}
\setcounter{equation}{0}
\renewcommand{\theequation}{\mbox{A.\arabic{equation}}}
In this appendix, we state the power counting rules for the chiral
quark model. Since we can treat the gluons perturbatively with
$\alpha_s\approx 0.28$\cite{georgi}, it suffices to focus on
Goldstone bosons and quarks.

The most general vertex in the chiral quark model in cutoff
regularization takes the form~\cite{georgi}, \ba
(2\pi)^4\delta^4(\sum p_{i})(\frac{\pi}{f_\pi})^A
(\frac{\psi}{f_\pi\sqrt{\Lambda}})^B
(\frac{gG_\mu}{\Lambda})^C(\frac{p}{\Lambda})^D f_\pi^2\Lambda^2
(\frac{m}{\Lambda})^{|\Delta \chi|/2}
 \ea
where $\Delta\chi$ is the chirality violation at the vertex, e.g.,
$\Delta\chi=2$ for each $m$. For notational simplicity, we write
$\Lambda$ for $\Lambda_{\chi SB}=4\pi f_\pi$.

Now as far as the Goldstone boson sector is concerned, the
counting rule is the same with the one given in standard chiral
perturbation theory(ChPT). Including quarks is straightforward
since the constituent quark mass can be considered small compared
to $\Lambda_{\chi SB}\sim 1$ GeV and hence $m_Q\sim p$ where $p$
is a typical momentum scale. Each quark propagator contributes
$-1$ power of $p$, each Goldstone boson propagator contributes
$-2$ power of $p$, each derivative and quark mass in the
interaction terms contribute $+1$ power of $p$ respectively and
each four momentum integration contributes $+4$ power of
$p$\footnote{Note that in the mass-independent substraction
scheme, the only dimensional parameter in the amplitude is the
momentum p.}.

Putting all the powers together, the chiral dimension $D$ of a
given amplitude with $L$ loops, $I_{GB}$ internal meson lines,
$I_{Q}$ quark lines, $N^{GB}$ mesonic vertices and $N^{GBQ}$
meson-quark vertices comes out to be
 \ba
D=4L -2I_{GB}-I_Q +\sum_n nN_n^{GB}+ \sum_d d N_d^{GBQ}.
 \ea
For connected diagrams, we can use the topological relation
 \ba
L=I_{GB}+I_Q - \sum_n (N_n^{GB}+N_n^{GBQ})+1
 \ea
to get \ba D=2L +2 +I_Q +\sum_n (n-2)N_n^{GB}+ \sum_d (d-2)
N_d^{GBQ}.
 \ea

\end{document}